\documentclass[11pt]{article}
\usepackage{cite}
\usepackage{graphicx}
\usepackage{indentfirst}
\usepackage{amssymb}
\usepackage{amsthm}
\usepackage{amsfonts}

\usepackage{anysize}
\usepackage{fancyhdr}
\begin{document}

\title{Rate Bounds for MIMO Relay Channels}
\author{Caleb K. Lo, Sriram Vishwanath and Robert W. Heath, Jr. \\Wireless Networking and Communications Group \\ Department of Electrical and Computer Engineering \\ The University of Texas at Austin \\ 1 University Station C0803 \\ Austin, TX 78712-0240 \\ Phone: (512) 471-1190 \\ Fax: (512) 471-6512 \\ Email: \{clo, sriram, rheath\}@ece.utexas.edu}
\date{}


\maketitle

\begin{abstract}
This paper considers the multi-input multi-output (MIMO) relay channel where multiple antennas are employed by each terminal.  Compared to single-input single-output (SISO) relay channels, MIMO relay channels introduce additional degrees of freedom, making the design and analysis of optimal cooperative strategies more complex.  In this paper, a partial cooperation strategy that combines transmit-side message splitting and block-Markov encoding is presented.  Lower bounds on capacity that improve on a previously proposed non-cooperative lower bound are derived for Gaussian MIMO relay channels\footnote{These lower bounds are special cases of \cite[Theorem 7]{CovGam:CapaTheoRelaChan:Sep:79}, but are of interest as they are applied to the MIMO relay channel and provide intuition about the structure of good coding strategies for MIMO relaying.}.
\end{abstract}

Keywords - Relay channels, MIMO systems, superposition coding, dirty-paper coding.

\section{Introduction}
\begin{figure}[tb]
\begin{center}
\includegraphics[width=3.0in]{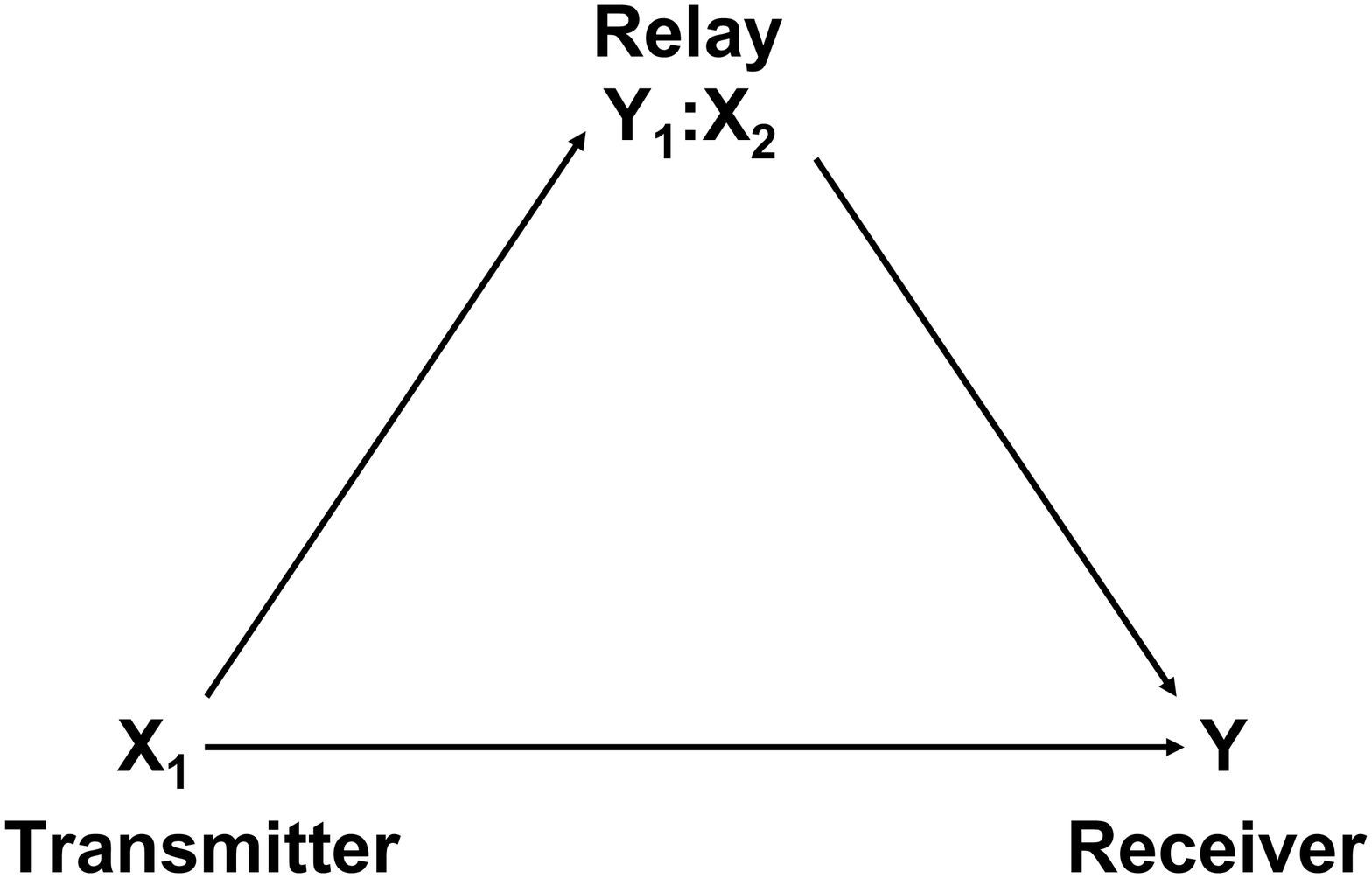}
\end{center}
\caption{Discrete memoryless relay channel.}
\label{relay-channel}
\end{figure}

Mesh networks that support multihop communication form an integral part of future-generation wireless communications \cite{PabETAL:RelaBaseDeplConc:Sep:04, BruETAL:MeshNetwCommMult:Mar:05, VisMuk:ThroRangTradWire:Mar:06}.  Relay channels are the fundamental building blocks of multihop mesh networks.  From \cite{CovGam:CapaTheoRelaChan:Sep:79}, a discrete memoryless relay channel is defined by $(\mathcal{X}_1\times\mathcal{X}_2,p(y,y_1|x_1,x_2),\mathcal{Y}\times\mathcal{Y}_1)$.  Here, $\mathcal{X}_1$, $\mathcal{X}_2$, $\mathcal{Y}_1$ and $\mathcal{Y}$ are finite sets corresponding to the transmitter, the relay and the receiver as shown in Fig. \ref{relay-channel}.

Relay channels were introduced in \cite{Meu:ThreTermCommChan:71} and upper bounds on their capacity were derived in \cite{Sat:InfoTranChanRela:Mar:76}.  Full-duplex relay channels were first analyzed from an information-theoretic perspective in \cite{CovGam:CapaTheoRelaChan:Sep:79}, where inner and outer bounds were derived and exact capacity expressions were obtained for special cases such as the physically degraded and Gaussian degraded relay channels.  The information-theoretic analysis in \cite{CovGam:CapaTheoRelaChan:Sep:79} relied on cooperation between the transmitter and the relay induced by block-Markov encoding.

Achievable rates in relay channels can be further improved via multi-input multi-output (MIMO) technology \cite{Fos:LayeArchWireComm:96,Win:CapaRadiCommSyst:Jun:87,MarHoc:CapaMobiCommLink:Jan:99, GesETAL:FromTheoPracOver:Apr:03}.  It has been shown that the capacity of a MIMO channel can scale linearly as the minimum of the number of transmit and receive antennas \cite{Tel:CapaMultGausChan:Nov:99}.  This encouraging result has led to research on multiuser MIMO channels such as Gaussian multiple access (MAC) \cite{CheVer:GausMultChanISI:May:93, Wyn:ShanApprGausCell:Nov:94, VisTseETAL:AsymOptiWateFill:Jan:01, YuRheETAL:IterGausVectMult:Jan:04} and broadcast (BC) \cite{CaiSha:AchiThroMultGaus:Jul:03, WeiSteETAL:CapaRegiGausMIMO:Sep:06, JinETAL:DualGausMultAcce:May:04, MohCio:ProoConvCapaGaus:Jul:06} channels.  Although discrete memoryless relay channels were analyzed decades ago, MIMO relay channels have only recently been studied \cite{WanETAL:CapaMIMORelaChan:Jan:05}.  As MIMO is an integral aspect of industry standards such as IEEE 802.16e \cite{WireMANWorkGrp}, and relaying is also being considered for practical implementation \cite{RelaTaskGrou}, it is natural to consider the performance limits of MIMO relaying.  In particular, MIMO relaying has gained increasing attention recently, and results have been obtained in terms of capacity scaling laws in large networks \cite{BolETAL:CapaScalLawsMIMO:Jun:06}, capacity scaling laws for two-way relaying \cite{VazHea:CapaScalMIMOTwo:Jun:07} and optimal precoder design \cite{TanHua:OptiDesiNonRege:Apr:07}.

In \cite{WanETAL:CapaMIMORelaChan:Jan:05}, a Gaussian relay channel with multiple antennas at each terminal is considered.  Upper and lower bounds on capacity are shown for both deterministic and Rayleigh fading channels.  The lower bounds for the case of fixed channels in \cite{WanETAL:CapaMIMORelaChan:Jan:05} arise from a non-cooperative transmit strategy.  Higher achievable rates than those yielded by the non-cooperative approach in \cite{WanETAL:CapaMIMORelaChan:Jan:05} can be obtained by observing that MIMO relay channels inherently contain more degrees of freedom than single-input single-output (SISO) relay channels, where each terminal employs only a single antenna.

We assume that the relay performs partial decode-and-forward operations, where the relay decodes a portion of the transmitter's codeword, encodes the decoded message using its own codebook, and sends the encoded message to the receiver.  In a MIMO relay channel, the channel eigenmodes can be exploited to optimize the cooperative role of the relay.  Thus, coding strategies such as transmit-side message splitting \cite{CovGam:CapaTheoRelaChan:Sep:79, GamAre:CapaSemiRelaChan:May:82} can increase the achievable rate for MIMO relay channels.

In Section V we consider a simple numerical example that illustrates the role that the channel eigenmodes play in optimizing the cooperative role of the relay for the MIMO case.  One scenario that we consider in Section V involves the transmitter-to-relay vector channel being orthogonal to the vector direct link.  In particular, this notion of orthogonality is a special case of the orthogonal relay channel considered in \cite{GamZah:CapaClasRelaChan:May:05}, where the general partial decode-and-forward strategy in \cite[Theorem 7]{CovGam:CapaTheoRelaChan:Sep:79} is shown to be capacity-achieving for both discrete-memoryless and Gaussian cases.  For the Gaussian case, the cooperative role of the relay is optimized by power allocation over both components of the transmitter's codeword, which is analogous to power allocation over the channel eigenmodes for the MIMO case.  

We present transmission strategies that rely on message splitting to support varying levels of cooperation between the transmitter and the relay in a MIMO relay channel.  In this policy, the transmitter has two messages and chooses its codeword as a function of both of them; the relay, though, only has to decode one of these messages.  The key intuition behind the application of message splitting to MIMO relaying is as follows: for the Gaussian SISO relay channel, while message splitting increases the average throughput \cite{GopETAL:SupeCodiBaseCoop:Oct:05, PopCar:SpecEffiWireRela:Apr:07}, it is not a capacity-achieving strategy.  This is based on the fact that in a Gaussian SISO relay channel, the received signals at the relay and the receiver can be statistically ordered.  On the other hand, the received signals in a Gaussian MIMO relay channel cannot be statistically ordered.  In particular, the channel eigenmodes determine the optimal level of cooperation between the transmitter and the relay in a MIMO relay channel, which is measured by how the transmitter chooses its codeword as a function of both messages.

We stress that our message splitting strategies are special cases of the partial cooperation approach in \cite[Theorem 7]{CovGam:CapaTheoRelaChan:Sep:79}.  Since a direct application of the general coding strategy in \cite[Theorem 7]{CovGam:CapaTheoRelaChan:Sep:79} to Gaussian MIMO relay channels would require a computationally intensive optimization over several auxiliary random variables, we consider simplified coding approaches and obtain closed-form achievable rate expressions.

We propose lower bounds on the capacity of the MIMO relay channel by utilizing transmit-side message splitting.  In particular, we consider both superposition coding and precoding at the transmitter.  For the case of precoding in a Gaussian MIMO relay channel, dirty-paper coding \cite{Cos:WritOnDirtPape:May:83} is employed at the transmitter.  Our proposed lower bounds obtained via a combination of transmit-side message splitting and block-Markov encoding improve on the lower bounds from \cite{WanETAL:CapaMIMORelaChan:Jan:05} that are obtained by a non-cooperative transmit strategy that does not employ block-Markov encoding.  The block-Markov encoding that we employ differs from the approach in \cite{CovGam:CapaTheoRelaChan:Sep:79} in that the relay cooperates with the transmitter over two consecutive transmission blocks to transmit only one of the transmitter's two messages.  The non-cooperative approach in \cite{WanETAL:CapaMIMORelaChan:Jan:05} is actually a special case of our proposed strategies.  We also perform a simple numerical analysis that illustrates how the achievable rate from our precoding approaches depends on the exact channel state and not just on the channel norms.

\begin{figure}[t]
\begin{center}
\includegraphics[width=3.0in]{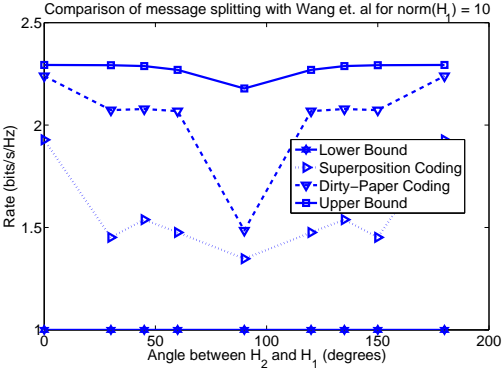}
\end{center}
\caption{Achievable rate results for the case where the transmitter, the relay, and the receiver are all equidistant from each other, or $\gamma_1$ = $\gamma_2$ = $\gamma_3$.}
\label{precode-wang-1}
\end{figure}

\begin{figure}[t]
\begin{center}
\includegraphics[width=3.0in]{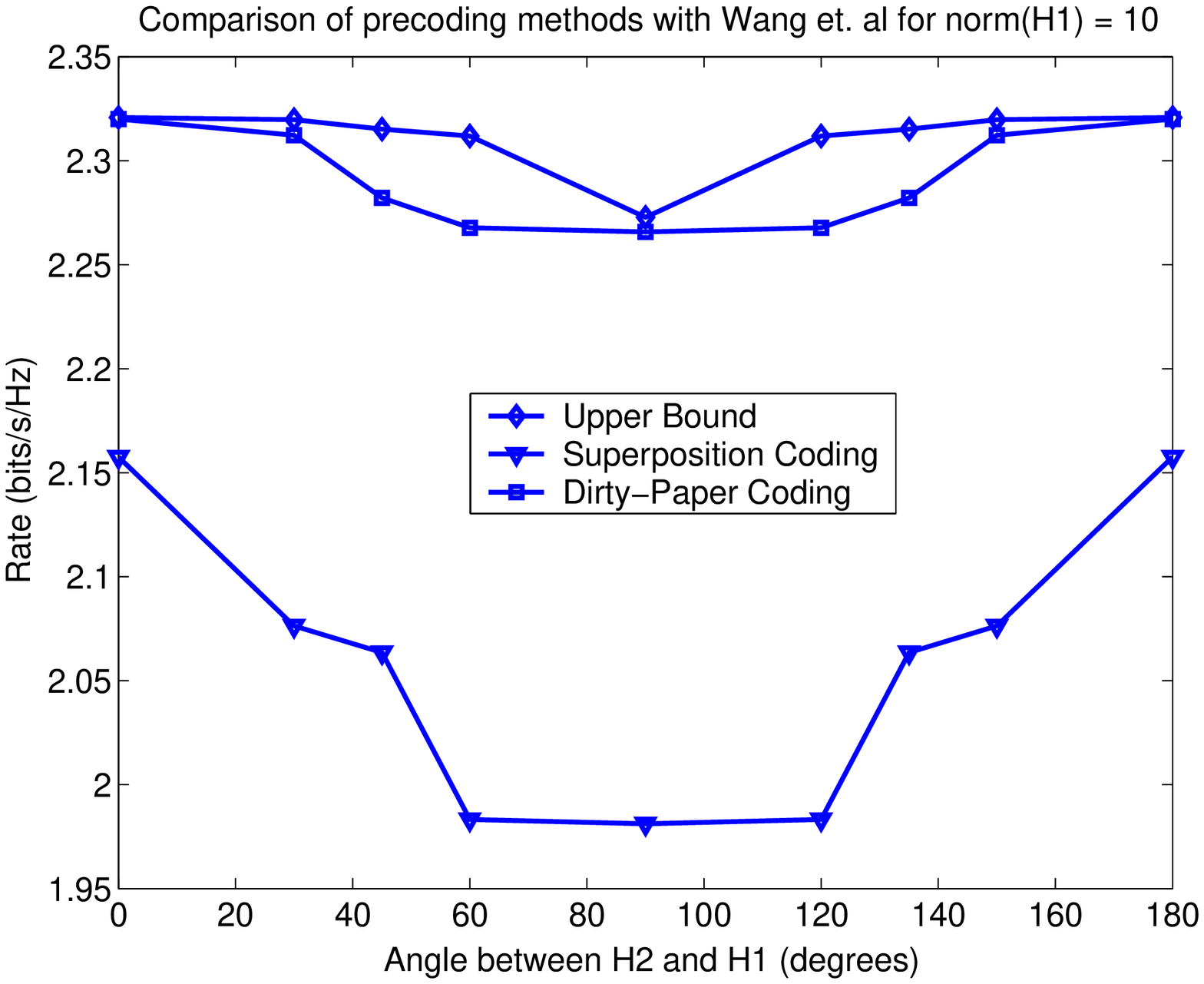}
\end{center}
\caption{Achievable rate results for the case where the relay is closer to the transmitter than to the receiver, or $\gamma_2$ = $\gamma_3$ and $\gamma_1$ = 10$\gamma_2$.}
\label{precode-wang-2}
\end{figure}

\begin{figure}[t]
\begin{center}
\includegraphics[width=3.0in]{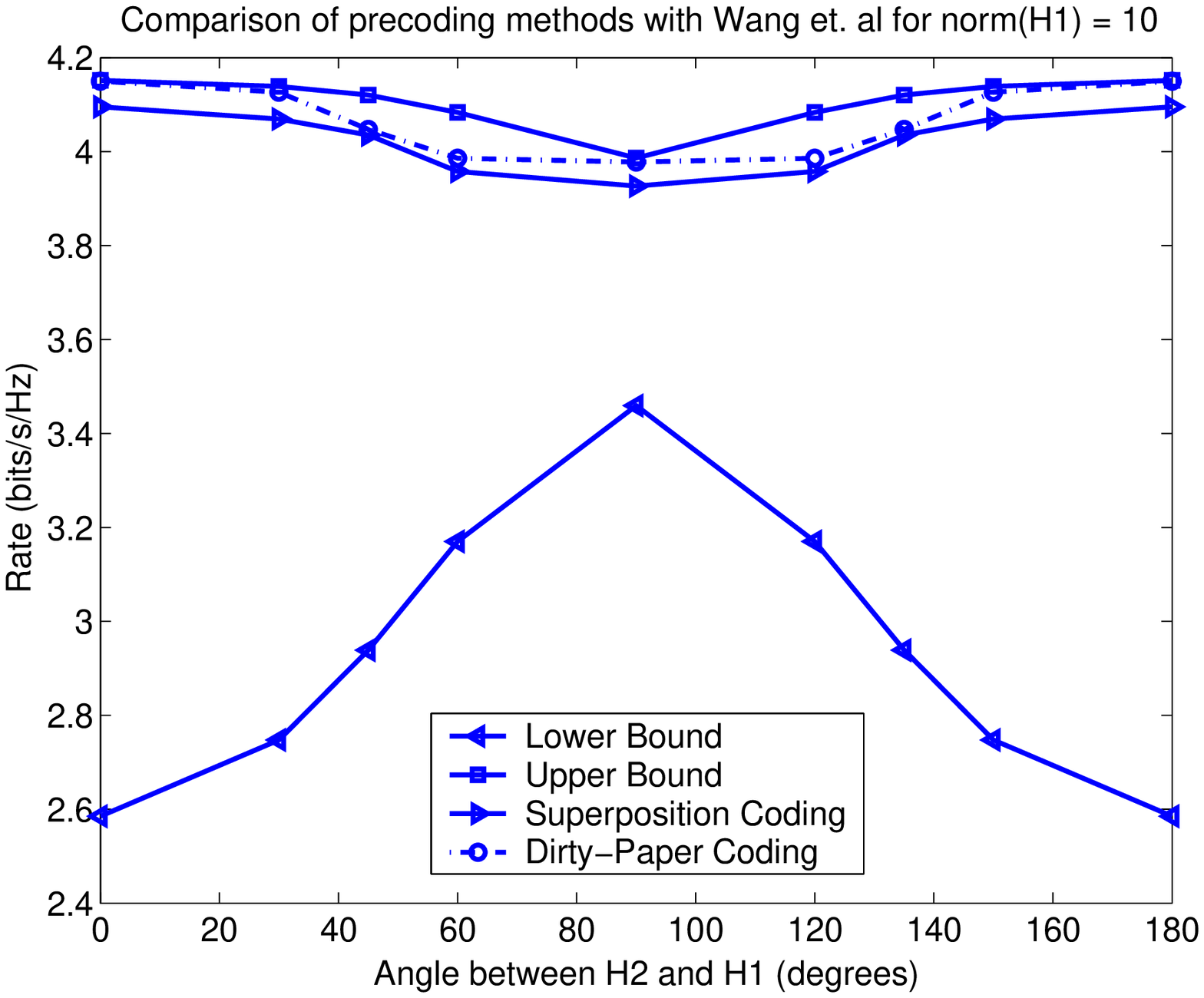}
\end{center}
\caption{Achievable rate results for the case where the relay is closer to the receiver than to the transmitter, or $\gamma_1$ = $\gamma_2$ and $\gamma_3$ = 10$\gamma_1$.}
\label{precode-wang-3}
\end{figure}

The rate bounds in this paper along with an initial version of the numerical results in Fig. \ref{precode-wang-1} were initially presented in \cite{LoETAL:RateBounMIMORela:Nov:05}.  This paper contains the full proofs of some of the key rate bounds, which lends valuable insights on the key encoding and decoding mechanisms for transmit-side message splitting in the MIMO relay channel.  We have also obtained revised numerical results for Fig. \ref{precode-wang-1}.  In addition, we have added Fig. \ref{precode-wang-2} and Fig. \ref{precode-wang-3}, which illustrate the impact of system topology on the derived rate bounds. 

This paper is organized as follows: In Section II we describe the system model.  Section III reviews the upper and lower bounds on capacity from \cite{WanETAL:CapaMIMORelaChan:Jan:05} for the Gaussian MIMO relay channel.  In Section IV, we present our message splitting strategies for Gaussian MIMO relay channels along with their associated achievable rates.  Numerical results are given in Section V.  We conclude the paper in Section VI.  The appendix contains rigorous derivations of some of the achievable rate expressions.  

We use boldface notation for matrices and vectors; uppercase notation is used for matrices while lowercase notation is used for vectors.  $\mathbb{E}$ represents mathematical expectation.  Re($x$) denotes the real part of a complex number $x$.  For a matrix $\textbf{A}$, $\textbf{A}^{\dagger}$, tr($\textbf{A}$) and det($\textbf{A}$) denote the transpose conjugate, trace, and determinant, respectively of $\textbf{A}$ while $\textbf{A}$ $\succeq$ 0 means that $\textbf{A}$ is positive semi-definite.  SNR represents signal-to-noise ratio.  $\textbf{I}_{K}$ denotes the $K$ $\times$ $K$ identity matrix.  We use $\mathcal{CN}$($\textbf{b}$, $\textbf{C}$) to represent the circularly symmetric complex Gaussian distribution with mean $\textbf{b}$ and covariance matrix $\textbf{C}$.  For a set $\mathcal{R}$, $\|\mathcal{R}\|$ denotes the cardinality of $\mathcal{R}$.

\section{System Model}\label{sysmod}
\begin{figure}[tb]
\begin{center}
\includegraphics[width=3.0in]{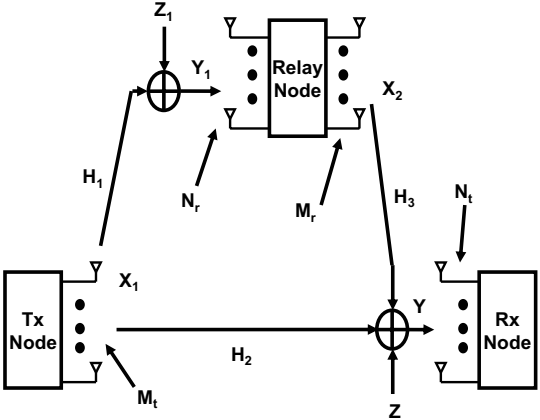}
\end{center}
\caption{Gaussian MIMO Relay Channel.}
\label{system-model}
\end{figure}

Consider the Gaussian MIMO full-duplex relay channel illustrated in Fig. \ref{system-model}.  Let $\textbf{x}_1$ and $\textbf{x}_2$ be the $M_t$ $\times$ 1 and $M_r$ $\times$ 1 transmitted signals from the transmitter and the relay.  Let $\textbf{y}$ and $\textbf{y}_1$ be the $N_t$ $\times$ 1 and $N_r$ $\times$ 1 received signals at the receiver and the relay.  Define $\textbf{H}_1$, $\textbf{H}_2$, and $\textbf{H}_3$ as $N_r$ $\times$ $M_t$, $N_t$ $\times$ $M_t$ and $N_t$ $\times$ $M_r$ channel gain matrices.  Define $\textbf{z}$ and $\textbf{z}_1$ as independent $N_t$ $\times$ 1 and $N_r$ $\times$ 1 circularly-symmetric complex Gaussian noise vectors with distributions $\mathcal{CN}$($\textbf{0}$, $\textbf{I}_{N_t}$) and $\mathcal{CN}$($\textbf{0}$, $\textbf{I}_{N_r}$).

We assume that the transmitter is subject to a power constraint $\mathbb{E}$($\textbf{x}_1^{\dagger}$$\textbf{x}_1$) $\leq$ $M_t$ and that the relay is also subject to a power constraint $\mathbb{E}$($\textbf{x}_2^{\dagger}$$\textbf{x}_2$) $\leq$ $M_r$.  We also assume that the relay has two sets of antennas, with one set for the receiver and one for the transmitter, so it operates in a full-duplex mode.  The relay also cancels out interference from its transmitter array at its receiver array.  In addition, we assume that all channel matrices are fixed and known at all three terminals and that $\textbf{z}$ and $\textbf{z}_1$ are uncorrelated with $\textbf{x}_1$ and $\textbf{x}_2$.  We do not consider fading channels in this paper.

We define parameters related to the SNR at the receiver and at the relay as $\gamma_1$ = $SNR_1$/$M_t$, $\gamma_2$ = $SNR_2$/$M_t$, and $\gamma_3$ = $SNR_3$/$M_r$ where $SNR_1$ and $SNR_2$ are the expected SNR values for $\textbf{x}_1$ after fading at each receive antenna at the relay and the receiver, and $SNR_3$ is the expected SNR for $\textbf{x}_2$ after fading at each receive antenna at the receiver \cite{MarHoc:CapaMobiCommLink:Jan:99}.

With these definitions, the received signals at the relay and at the receiver are
\begin{equation}\label{sysmodeq}
\begin{array}{lll}
\textbf{y}_1 & = & \sqrt{\gamma_1} \textbf{H}_1 \textbf{x}_1 + \textbf{z}_1 \\
\textbf{y} & = & \sqrt{\gamma_2} \textbf{H}_2 \textbf{x}_1 + \sqrt{\gamma_3} \textbf{H}_3 \textbf{x}_2 + \textbf{z}.
\end{array}
\end{equation}

\subsection{Weak Typicality}
Our proofs in this paper rely on the notion of weak typicality \cite{CovTho:ElemInfoTheo:06}.  Let $X\sim p(x)$ be a random variable.  The set $A_{\epsilon}^{(n)}(X)$ of weakly typical sequences $x^n$, where $p(x^n) = \prod_{i=1}^np(x_i)$ is
\begin{eqnarray}
A_{\epsilon}^{(n)}(X) & = & \bigg\{x^n: \bigg|-\frac{1}{n}\log p(x^n) - H(X)\bigg| < \epsilon\bigg\}. \nonumber
\end{eqnarray}

Note that our results for discrete memoryless channels based on finite alphabet codebooks can be generalized in a straightforward manner to Gaussian channels with Gaussian codebooks.  This generalization is based on applying weak typicality to continuous distributions that are subject to second moment constraints \cite{CovTho:ElemInfoTheo:06}.

\section{Background}\label{bckgrnd}
It was shown in \cite[Sec. III]{CovGam:CapaTheoRelaChan:Sep:79} that the capacity $C$ of a general relay channel is upper-bounded as
\begin{equation}\label{GenBound}
C \leq \max_{p(x_1, x_2)} \min\{I(X_1;Y,Y_1|X_2), I(X_1,X_2;Y)\}
\end{equation}
where the first term in the minimization is the rate from the transmitter to the relay and the receiver and the second term is the rate from the transmitter and the relay to the receiver.

Now let $\textbf{x}_1$ and $\textbf{x}_2$ be random vectors with mean zero and covariance matrices $\bf{\Sigma}$$_{ij}$ = $\mathbb{E}$($\textbf{x}_i$$\textbf{x}_j^{\dagger}$).  The authors of \cite{WanETAL:CapaMIMORelaChan:Jan:05} established the following capacity upper bound and lower bound for the case where the channel gains are fixed and known at each terminal.

\newtheorem{thm31}{Lemma}[section]
\begin{thm31}\label{thm31}\textnormal{\cite[Sec. III]{WanETAL:CapaMIMORelaChan:Jan:05}}
An upper bound on the capacity of the Gaussian MIMO relay channel is given by
\begin{equation}
C^G \leq C_{upper}^G = \max_{0\leq\rho\leq 1, {\bf{\Sigma}}_{11}, {\bf{\Sigma}}_{22}} \min(C_1^G, C_2^G)
\end{equation}
where tr($\bf{\Sigma}$$_{11}$) $\leq$ $M_t$, tr($\bf{\Sigma}$$_{22}$) $\leq$ $M_r$ and
\begin{equation}
\begin{array}{lll}
C_1^G & \triangleq & \log\left[\det\left(\textbf{I}_{M_t} + (1-\rho^2) \left[ \matrix{\sqrt{\gamma_1}\textbf{H}_1 \cr \sqrt{\gamma_2}\textbf{H}_2 \cr} \right] {\bf{\Sigma}}_{11} \left[ \matrix{\sqrt{\gamma_1}\textbf{H}_1 \cr \sqrt{\gamma_2}\textbf{H}_2 \cr} \right]^{\dagger}\right)\right] \\
C_2^G & \triangleq & \inf_{a>0}\log[\det(\textbf{I}_{N_t} + (\gamma_2 + \frac{\rho^2}{a}\sqrt{\gamma_2\gamma_3})\textbf{H}_2{\bf{\Sigma}}_{11}\textbf{H}_2^{\dagger} + (\gamma_3+a\sqrt{\gamma_2\gamma_3})\textbf{H}_3{\bf{\Sigma}}_{22}\textbf{H}_3^{\dagger})].
\end{array}
\end{equation}
\end{thm31}
As stated in \cite[Sec. IIIA]{WanETAL:CapaMIMORelaChan:Jan:05}, $\rho$ represents the correlation between $\textbf{x}_1$ and $\textbf{x}_2$.  Also, $a$ is a constant that arises from the vector-valued inequality in \cite[Lemma 3.2]{WanETAL:CapaMIMORelaChan:Jan:05}.  In addition, $C_1^G$ and $C_2^G$ represent the maximum sum rate across the transmitter-side broadcast cut and receiver-side multiple-access cut, respectively, in the Gaussian MIMO relay channel.

\newtheorem{thm32}[thm31]{Lemma}
\begin{thm32}\label{thm32}\textnormal{\cite[Sec. III]{WanETAL:CapaMIMORelaChan:Jan:05}}
A lower bound on the capacity of the Gaussian MIMO relay channel is given by
\begin{equation}
C^G \geq C_{lower}^G = \max(C_d^G, \min(C_3^G, C_4^G)),
\end{equation}
where
\begin{equation}\label{C4GDef}
\begin{array}{lll}
C_d^G & \triangleq & \max_{{\bf{\Sigma}}_{11}}\log[\det(\textbf{I}_{N_t} + \gamma_2\textbf{H}_2{\bf{\Sigma}}_{11}\textbf{H}_2^{\dagger})] \\
C_3^G & \triangleq & \max_{{\bf{\Sigma}}_{11}}\log[\det(\textbf{I}_{N_r} + \gamma_1\textbf{H}_1{\bf{\Sigma}}_{11}\textbf{H}_1^{\dagger})] \\
C_4^G & \triangleq & \max_{{\bf{\Sigma}}_{22}}\log[\det(\textbf{I}_{N_t} + \gamma_3\textbf{H}_3{\bf{\Sigma}}_{22}\textbf{H}_3^{\dagger}(\textbf{I}_{N_t} + \gamma_2\textbf{H}_2{\bf{\Sigma}}_{11}^{*}\textbf{H}_2^{\dagger})^{-1})]
\end{array}
\end{equation}
with
\begin{equation}
{\bf{\Sigma}}_{11}^{*} \triangleq \arg \max_{{\bf{\Sigma}}_{11} \succeq 0}\log[\det(\textbf{I}_{N_r} + \gamma_1\textbf{H}_1{\bf{\Sigma}}_{11}\textbf{H}_1^{\dagger})].
\end{equation}
\end{thm32}

Our objective is to use transmit-side message splitting to improve upon the bound in Lemma \ref{thm32}.  We outline this strategy in the next section.

\section{Transmit-Side Message Splitting}
Next we describe the transmission strategy that is employed in this paper.  We divide the transmit message into two components, denoted by the random variables $w_u$ and $w_v$.  $w_u$ is the message that is decoded by the relay and is thus cooperatively sent by the transmitter-relay pair to the receiver.  $w_v$, however, is intended to be decoded only by the receiver, and thus is a source of ``interference'' at the relay that is known a-priori at the transmitter.

We consider two classes of transmission strategies with this setup.  The first is superposition coding, where codebooks for $w_u$ and $w_v$ are determined separately and then simply superposed (added to one another) at the transmitter.  The second strategy is to utilize precoding at the transmit end, where intuitively the transmitter attempts to mitigate the interference caused by $w_v$ to the desirable signal corresponding to $w_u$ at the relay.  For both strategies, the transmitter and the relay cooperate in block-Markov encoding of $w_u$.

Note that the receiver must determine both $w_u$ and $w_v$ to decode the transmit message.  Thus, if $R_u$ denotes the rate for the codebook corresponding to $w_u$ and $R_v$ that for $w_v$, the net achievable rate for both superposition coding and precoding is $R = R_u + R_v$.  Assuming the receiver successively decodes $w_u$ and $w_v$, the order in which they are decoded impacts their rates.  In this paper, we use both decoding orders and choose the order that maximizes the overall rate.

Let $\textbf{u}$ and $\textbf{v}$ be auxiliary variables representing the contribution of $w_u$ and $w_v$, respectively to $\textbf{x}_1$.  Define ${\bf{\Sigma}}_u$, ${\bf{\Sigma}}_v$ and ${\bf{\Sigma}}_{x_2}$ to be the covariance matrices of $\textbf{u}$, $\textbf{v}$ and $\textbf{x}_2$ respectively.  Also, define \[\textbf{A} = \left[ \matrix{{\bf{\Sigma}}_u & \mathbb{E}(\textbf{u}\textbf{x}_2^{\dagger}) \cr \mathbb{E}(\textbf{x}_2\textbf{u}^{\dagger}) & {\bf{\Sigma}}_{x_2} \cr} \right]\] and $\textbf{B}$ = $\left[ \matrix{\sqrt{\gamma_2}\textbf{H}_2 & \sqrt{\gamma_3}\textbf{H}_3 \cr } \right]$.  In this case, $\mathbb{E}$($\textbf{u}$$\textbf{u}^{\dagger}$) = ${\bf{\Sigma}}_u$.  In addition, define $\mathcal{X}_1$, $\mathcal{X}_2$, $\mathcal{U}$ and $\mathcal{V}$ as the finite alphabets for $\textbf{x}_1$, $\textbf{x}_2$, $\textbf{u}$ and $\textbf{v}$, respectively.

\subsection{Superposition Coding}
\begin{figure}[tb]
\begin{center}
\includegraphics[width=3.0in]{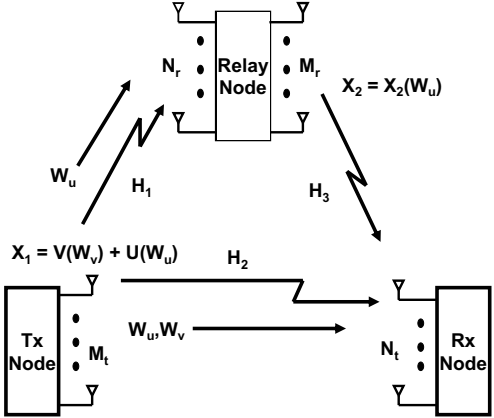}
\end{center}
\caption{Gaussian MIMO relay channel with superposition coding.}
\label{superposition-coding}
\end{figure}

Consider the system illustrated in Fig. \ref{superposition-coding}.  Assume that the receiver attempts to decode $w_u$ before decoding $w_v$.  Let $R_{sc,u}$ be the achievable rate for this case.  By applying the partial cooperation strategy of \cite[Theorem 7]{CovGam:CapaTheoRelaChan:Sep:79} to this case, it can be proved that
\begin{equation}\label{rate-sup-s}
R_{sc,u} = \sup_{p(x_1,x_2,u,v)} (R_{sc,u,1} + R_{sc,u,2})
\end{equation}
where
\begin{equation}
\begin{array}{lll}
R_{sc,u,1} & = & \min(I(U;Y_1|X_2),I(U,X_2;Y)) \\
R_{sc,u,2} & = & I(V;Y|U,X_2)
\end{array}
\end{equation}
and the supremum is taken over all joint distributions \[p(x_1,x_2,u,v) = p(x_2)p(u|x_2)p(v)p(x_1|u,v)\] on $\mathcal{X}_1\times\mathcal{X}_2\times\mathcal{U}\times\mathcal{V}$.  In particular, $I(U;Y_1|X_2)$ is the maximum signaling rate for $w_u$ over the transmitter-to-relay link.  Also, $I(U,X_2;Y)$ is the maximum signaling rate for $w_u$ over the effective multiple-access channel from the transmitter and relay to the receiver.  In addition, $I(V;Y|U,X_2)$ is the maximum signaling rate for $w_v$ over the transmitter-to-receiver link.

For the Gaussian MIMO relay channel, we employ Gaussian codebooks for $\textbf{u}$ and $\textbf{v}$ at the transmitter.  We prove in Appendix \ref{proofstr} that
\begin{equation}\label{str}
I(U;Y_1|X_2) = \log\left(\frac{\det\left(\textbf{I}_{N_r}+\gamma_1\textbf{H}_1\left({\bf{\Sigma}}_u-\mathbb{E}(\textbf{u}\textbf{x}_2^{\dagger}){\bf{\Sigma}}_{x_2}^{-1}\mathbb{E}(\textbf{x}_2\textbf{u}^{\dagger})+{\bf{\Sigma}}_v\right)\textbf{H}_1^{\dagger}\right)}{\det\left(\textbf{I}_{N_r}+\gamma_1\textbf{H}_1{\bf{\Sigma}}_v\textbf{H}_1^{\dagger}\right)}\right),
\end{equation}
\begin{equation}\label{smac}
I(U,X_2;Y) = \log\left(\frac{\det\left(\textbf{I}_{N_t}+\gamma_2\textbf{H}_2{\bf{\Sigma}}_v\textbf{H}_2^{\dagger}+\textbf{B}\textbf{A}\textbf{B}^{\dagger}\right)}{\det\left(\textbf{I}_{N_t}+\gamma_2\textbf{H}_2{\bf{\Sigma}}_v\textbf{H}_2^{\dagger}\right)}\right),
\end{equation}
and
\begin{equation}\label{ctr}
I(V;Y|U,X_2) = \log(\det(\textbf{I}_{N_t}+\gamma_2\textbf{H}_2{\bf{\Sigma}}_v\textbf{H}_2^{\dagger})).
\end{equation} 

Now assume that the receiver attempts to decode $w_v$ before decoding $w_u$.  Let $R_{sc,v}$ be the achievable rate for this case.  By applying the partial cooperation strategy of \cite[Theorem 7]{CovGam:CapaTheoRelaChan:Sep:79} to this case, it can be proved that
\begin{equation}\label{rate-sup-c}
R_{sc,v} = \sup_{p(x_1,x_2,u,v)} (R_{sc,v,1} + R_{sc,v,2})
\end{equation}
where
\begin{equation}
\begin{array}{lll}
R_{sc,v,1} & = & \min(I(U;Y_1|X_2),I(U,X_2;Y|V)) \\
R_{sc,v,2} & = & I(V;Y)
\end{array}
\end{equation}
and the supremum is taken over all joint distributions \[p(x_1,x_2,u,v) = p(x_2)p(u|x_2)p(v)p(x_1|u,v)\] on $\mathcal{X}_1\times\mathcal{X}_2\times\mathcal{U}\times\mathcal{V}$.

In this case our choice of Gaussian codebooks for $\textbf{u}$ and $\textbf{v}$ in a Gaussian MIMO relay channel yields
\begin{equation}\label{ctr2}
I(V;Y) = \log\left(\frac{\det\left(\textbf{I}_{N_t}+\gamma_2\textbf{H}_2{\bf{\Sigma}}_v\textbf{H}_2^{\dagger}+\textbf{B}\textbf{A}\textbf{B}^{\dagger}\right)}{\det\left(\textbf{I}_{N_t}+\textbf{B}\textbf{A}\textbf{B}^{\dagger}\right)}\right),
\end{equation}
which is analogous to the rate in (\ref{ctr}) and
\begin{equation}\label{smac2}
I(U,X_2;Y|V) = \log(\det(\textbf{I}_{N_t}+\textbf{B}\textbf{A}\textbf{B}^{\dagger})),
\end{equation}
which is analogous to the rate in (\ref{smac}) while I($U;Y_1|X_2$) is the same as in (\ref{str}).

The objective is to choose the decoding order that yields a higher overall rate.  We now state the following intuitively obvious result, which will not require a formal proof.

\newtheorem{thm41}{Proposition}[section]
\begin{thm41}\label{thm41}
Let $R_{sc}$ be the maximum signaling rate for the Gaussian MIMO relay channel where the transmitter employs superposition coding.  Then
\begin{equation}
R_{sc} = \max(R_{sc,u}, R_{sc,v}) \geq C_{lower}^G
\end{equation}
where $C_{lower}^G$ is given in Lemma \ref{thm32}.
\end{thm41}

By setting $\textbf{v}$ = $\textbf{x}_1$ and $\textbf{u}$ = 0, we can achieve $C_d^G$.  Also, by setting $\textbf{u}$ = $\textbf{x}_1$, $\textbf{v}$ = 0 and having the relay employ a codebook of the same cardinality as that of the codebook at the transmitter, we can achieve at least $\min(C_3^G, C_4^G)$.

\subsection{Precoding}
Instead of superposition coding, consider a strategy where the transmitter uses precoding to mitigate the interference caused by $w_v$ to the desired signal corresponding to $w_u$ at the relay.  Assume that the receiver attempts to decode $w_u$ before decoding $w_v$.  Let $R_{pre,u}$ be the achievable rate for this case.  It is proved in Appendix \ref{rdpcs} that
\begin{equation}\label{rate-pre-s}
R_{pre,u} = \sup_{p(x_1,x_2,u,v)} (R_{pre,u,1} + R_{pre,u,2})
\end{equation}
where
\begin{equation}
\begin{array}{lll}
R_{pre,u,1} & = & \min(I(U;Y_1|X_2) - I(U;V|X_2), I(U,X_2;Y)) \\
R_{pre,u,2} & = & I(V;Y|U,X_2)
\end{array}
\end{equation}
and the supremum is taken over all joint distributions \[p(x_1,x_2,u,v) = p(v)p(x_2)p(u|x_2)p(x_1|u,v)\] on $\mathcal{X}_1\times\mathcal{X}_2\times\mathcal{U}\times\mathcal{V}$.  Note from the form of the joint distributions that $\textbf{u}$ and $\textbf{v}$ are correlated, which differs from the case of superposition coding.  The transmitter selects $\textbf{u}$ as a function of the known interference $\textbf{v}$ on the transmitter-to-relay channel $\textbf{H}_1$.

For the Gaussian MIMO relay channel, we employ Gaussian codebooks for $\textbf{u}$ and $\textbf{v}$.  In particular, we choose $\textbf{u}$ = $\textbf{G}\textbf{v}$ + $\textbf{x}_1^{'}$ and $\textbf{x}_1$ = $\textbf{x}_1^{'}$ + $\textbf{v}$, where $\textbf{x}_1^{'}$ and $\textbf{v}$ are chosen to be independent.  Thus we are employing dirty-paper coding at the transmitter and the objective is to choose $\textbf{G}$ to maximize $I(U;Y_1|X_2)$ - $I(U;V|X_2)$.  We define ${\bf{\Sigma}}_{x_1^{'}|x_2}$ to be the covariance matrix of $\textbf{x}_1^{'}$ given knowledge of $\textbf{x}_2$.  By following a procedure similar to that in \cite[Appendix C]{Yu:CompCoopMultUser:Jun:02}, we have
\begin{equation}\label{str2}
I(U;Y_1|X_2) - I(U;V|X_2) = \log(\det(\textbf{I}_{N_r}+\gamma_1\textbf{H}_1{\bf{\Sigma}}_{x_1^{'}|x_2}\textbf{H}_1^{\dagger})),
\end{equation}
which is analogous to the rate in (\ref{str}); I($U,X_2;Y$) and I($V;Y|U,X_2$) are the same as in (\ref{smac}) and (\ref{ctr}) respectively.

Now assume that the receiver attempts to decode $w_v$ before decoding $w_u$.  Let $R_{pre,v}$ be the achievable rate for this case.  It is proved in Appendix \ref{rdpcc} that
\begin{equation}\label{rate-pre-c}
R_{pre,v} = \sup_{p(x_1,x_2,u,v)} (R_{pre,v,1} + R_{pre,v,2})
\end{equation}
where
\begin{equation}
\begin{array}{lll}
R_{pre,v,1} & = & \min(I(U;Y_1|X_2) - I(U;V|X_2), I(U,X_2;Y|V)) \\
R_{pre,v,2} & = & I(V;Y)
\end{array}
\end{equation}
and the supremum is taken over all joint distributions \[p(x_1,x_2,u,v) = p(x_2)p(u,v|x_2)p(x_1|u,v)\] on $\mathcal{X}_1$ $\times$ $\mathcal{X}_2$ $\times$ $\mathcal{U}$ $\times$ $\mathcal{V}$.

In this case our choice of dirty-paper coding at the transmitter in a Gaussian MIMO relay channel results in I($V;Y$), I($U;Y_1|X_2$) $-$ I($U;V|X_2$) and I($U,X_2;Y|V$) being the same as in (\ref{ctr2}), (\ref{str2}), and (\ref{smac2}) respectively.

The objective is to choose the decoding order that yields a higher overall rate.  We now state and prove the following result.

\newtheorem{thm42}[thm41]{Proposition}
\begin{thm42}\label{thm42}
Let $R_{pre}$ be the maximum signaling rate for the Gaussian MIMO relay channel employing dirty-paper coding at the transmitter.  Then
\begin{equation}
R_{pre} = \max(R_{pre,u}, R_{pre,v}) \geq R_{sc}
\end{equation}
where $R_{sc}$ is given in Proposition \ref{thm41}.
\end{thm42}

\begin{proof}
We show that superposition coding is a special case of our precoding strategy.  Without loss of generality, assume that $w_u$ is decoded first at the receiver.  Recall that
\begin{eqnarray}
R_{pre,u} & = & \sup_{p(v)p(x_2)p(u|v,x_2)p(x_1|u,v)} (\min(I(U;Y_1|X_2) - I(U;V|X_2), \label{pre-expr} \\
& & I(U,X_2;Y)) + I(V;Y|U,X_2)). \nonumber
\end{eqnarray}
By considering the case where $\textbf{u}$ and $\textbf{v}$ are independent random variables, we find that $p(u|v,x_2) = p(u|x_2)$ and $I(U;V|X_2) = 0$.  Thus, (\ref{pre-expr}) reduces to
\begin{equation}
R_{sc,u} = \sup_{p(x_2)p(u|x_2)p(v)p(x_1|u,v)} (\min(I(U;Y_1|X_2),I(U,X_2;Y)) + I(V;Y|U,X_2)).
\end{equation}

It immediately follows that $R_{pre}$ $\geq$ $R_{sc}$.
\end{proof}

\section{Numerical Results}
We employ a simple example to demonstrate how transmit-side message splitting outperforms the bounds in \cite[Sec. III]{WanETAL:CapaMIMORelaChan:Jan:05}.  We choose $\textbf{H}_2 = [1~0]$ and $\textbf{H}_3 = 1$.  We also choose $\textbf{H}_1 = [x~y]$, where $x,y\in\mathbb{R}$, and constrain $\|\textbf{H}_1\| = 10$.  By considering $\textbf{H}_1$ and $\textbf{H}_2$ as two-dimensional vectors, we can define an ``angle'' $\Theta(\textbf{H}_1,\textbf{H}_2)$ between them.  We vary $\Theta(\textbf{H}_1, \textbf{H}_2)$ over the range $[0,\pi]$, where $\Theta(\textbf{H}_1,\textbf{H}_2)$ is expressed in radians.  As $\Theta(\textbf{H}_1,\textbf{H}_2)\rightarrow\pi/2$, the gain between the second transmit antenna and the relay's antenna, or $y$, increases.  Note that the norm constraint on $\textbf{H}_1$ causes the gain between the first transmit antenna and the relay's antenna, or $x$, to decrease.

We consider three system topologies.  The first topology is where the transmitter, the relay, and the receiver are equidistant, and this is modeled by setting $\gamma_1 = \gamma_2 = \gamma_3$ in (\ref{sysmodeq}).  The second topology is where the relay is closer to the transmitter than to the receiver, and this is modeled by setting $\gamma_2 = \gamma_3$ and $\gamma_1 = 10\gamma_2$ in (\ref{sysmodeq}).  The third topology is where the relay is closer to the receiver than to the transmitter, and this is modeled by setting $\gamma_1 = \gamma_2$ and $\gamma_3 = 10\gamma_1$ in (\ref{sysmodeq}).  For all three topologies, we observed that the lower bound from \cite[Sec. III]{WanETAL:CapaMIMORelaChan:Jan:05} is 1 bits/s/Hz for all values of $\Theta(\textbf{H}_1,\textbf{H}_2)$, which results from our fixing $\textbf{H}_2$ at $[1~0]$.

We need to solve the optimization problems (\ref{rate-sup-s}), (\ref{rate-sup-c}), (\ref{rate-pre-s}) and (\ref{rate-pre-c}) to obtain the achieved rates for our message splitting strategies.  We employ numerical direct search methods such as the Nelder-Mead algorithm, differential evolution and simulated annealing to solve (\ref{rate-sup-s}), (\ref{rate-sup-c}), (\ref{rate-pre-s}) and (\ref{rate-pre-c}).

Fig. \ref{precode-wang-1} shows the rates that are achieved by our message splitting strategies along with the upper and lower bounds from \cite[Sec. III]{WanETAL:CapaMIMORelaChan:Jan:05} for the first topology.  We see that the upper bound decreases as $\Theta(\textbf{H}_1,\textbf{H}_2)\rightarrow\pi/2$ radians.  Also, as $\Theta(\textbf{H}_1,\textbf{H}_2)\rightarrow\pi/2$, the transmitter uses more power on its second transmit antenna to exploit the rate benefits on the transmitter-to-relay link.  This strategy, though, results in a loss of rate on the direct link since $\textbf{H}_2$ is fixed at $[1~0]$.  This leads to a monotonic decrease in the upper bound as $\Theta(\textbf{H}_1,\textbf{H}_2)\rightarrow\pi/2$.

We see that the achievable rates via superposition coding and dirty-paper coding always outperform the lower bound of 1 bits/s/Hz.  Also, we see that the achievable rate from dirty-paper coding is never less than the achievable rate from superposition coding.

Fig. \ref{precode-wang-2} shows the rates that are achieved by our transmit-side message splitting strategies along with the upper bound from \cite[Sec. III]{WanETAL:CapaMIMORelaChan:Jan:05} for the second topology.  As in Fig. \ref{precode-wang-1}, we see that the upper bound and our achievable rates via message splitting monotonically decrease as $\Theta(\textbf{H}_1, \textbf{H}_2)$ $\rightarrow$ $\pi$/2 radians.

Fig. \ref{precode-wang-3} shows the rates that are achieved by our transmit-side message splitting strategies along with both the upper and lower bounds from \cite[Sec. III]{WanETAL:CapaMIMORelaChan:Jan:05} for the third topology.  Here, we see that the lower bound monotonically increases as $\Theta(\textbf{H}_1, \textbf{H}_2)$ $\rightarrow$ $\pi$/2 radians.  Also, superposition coding performs comparably to dirty-paper coding over all angles in this case, whereas for the other two topologies dirty-paper coding generally significantly outperformed superposition coding.

\section{Conclusion}
We derived new lower bounds on capacity for MIMO relay channels via transmit-side message splitting.  Our proposed bounds improve upon the lower bounds that were introduced in \cite{WanETAL:CapaMIMORelaChan:Jan:05}.  In particular, our results show the benefits of employing the relay's assistance via superposition coding and precoding at the transmitter.  Our results suggest that transmit-side message splitting should be an integral part of communication over MIMO relay channels, especially when the transmitter-to-relay link is strong relative to the transmitter-to-receiver and/or relay-to-receiver channels.


\appendix

\section{Proofs Of Rate Bounds}

\subsection{Establishing (\ref{str})}\label{proofstr}
We have $I(U;Y_1|X_2)$ = $h(Y_1|X_2)$ - $h(Y_1|X_2,U)$.  Since the transmitter employs superposition coding, we have
\begin{equation}
\begin{array}{lll}
\textbf{y}_1 & = & \sqrt{\gamma_1}\textbf{H}_1\textbf{x}_1 + \textbf{z}_1 \\
& = & \sqrt{\gamma_1}\textbf{H}_1(\textbf{u} + \textbf{v}) + \textbf{z}_1
\end{array}
\end{equation}
and since $\textbf{u}$ and $\textbf{v}$ are independent given $\textbf{x}_2$, and $\textbf{v}$ and $\textbf{x}_2$ are independent, we have
\begin{equation}
\begin{array}{lll}
h(Y_1|X_2) & = & h(\sqrt{\gamma_1}\textbf{H}_1(U+V)+Z_1|X_2) \\
& = & \log((2\pi e)^{N_r}\det(\gamma_1\textbf{H}_1({\bf{\Sigma}}_{u|x_2}+{\bf{\Sigma}}_{v})\textbf{H}_1^{\dagger}+\textbf{I}_{N_r}))
\end{array}
\end{equation}
and since $\textbf{z}_1$ is independent of $\textbf{u}$, $\textbf{v}$ and $\textbf{x}_2$ we have
\begin{equation}
\begin{array}{lll}
h(Y_1|X_2,U) & = & h(\sqrt{\gamma_1}\textbf{H}_1(U+V)+Z_1|X_2,U) \\
& = & h(\sqrt{\gamma_1}\textbf{H}_1V+Z_1|X_2,U) \\
& = & h(\sqrt{\gamma_1}\textbf{H}_1V+Z_1) \\
& = & \log((2\pi e)^{N_r}\det(\gamma_1\textbf{H}_1{\bf{\Sigma}}_{v}\textbf{H}_1^{\dagger}+\textbf{I}_{N_r})).
\end{array}
\end{equation}

Now we note that $\log((2\pi e)^{M_t}\det(\bf{\Sigma}$$_{u|x_2}))$ = $h(U|X_2)$ = $h(U,X_2)$ - $h(X_2)$ = \\
$\log((2\pi e)^{M_t+M_r}\det(A))$ - $\log((2\pi e)^{M_r}\det(\bf{\Sigma}$$_{x_2}))$ where \[ A = \left[ \matrix{{\bf{\Sigma}}_u & \mathbb{E}(\textbf{u}\textbf{x}_2^{\dagger}) \cr \mathbb{E}(\textbf{x}_2\textbf{u}^{\dagger}) & {\bf{\Sigma}}_{x_2} \cr} \right] \] so \[\det(A) = \det({\bf{\Sigma}}_{x_2}) \cdot \det({\bf{\Sigma}}_u-\mathbb{E}(\textbf{u}\textbf{x}_2^{\dagger}){\bf{\Sigma}}_{x_2}^{-1}\mathbb{E}(\textbf{x}_2\textbf{u}^{\dagger}))\] and \[{\bf{\Sigma}}_{u|x_2} = {\bf{\Sigma}}_u - \mathbb{E}(\textbf{u}\textbf{x}_2^{\dagger}){\bf{\Sigma}}_{x_2}^{-1}\mathbb{E}(\textbf{x}_2\textbf{u}^{\dagger}).\]

Thus we have
\begin{equation}
h(Y_1|X_2) = \log((2\pi e)^{N_r}\det(\gamma_1\textbf{H}_1({\bf{\Sigma}}_u - \mathbb{E}(\textbf{u}\textbf{x}_2^{\dagger}){\bf{\Sigma}}_{x_2}^{-1}\mathbb{E}(\textbf{x}_2\textbf{u}^{\dagger})+{\bf{\Sigma}}_v)\textbf{H}_1^{\dagger}+\textbf{I}_{N_r}))
\end{equation}
and finally we obtain
\begin{equation}
I(U;Y_1|X_2) = \log\left(\frac{\det\left(\textbf{I}_{N_r}+\gamma_1\textbf{H}_1\left({\bf{\Sigma}}_u-\mathbb{E}(\textbf{u}\textbf{x}_2^{\dagger}){\bf{\Sigma}}_{x_2}^{-1}\mathbb{E}(\textbf{x}_2\textbf{u}^{\dagger})+{\bf{\Sigma}}_v\right)\textbf{H}_1^{\dagger}\right)}{\det\left(\textbf{I}_{N_r}+\gamma_1\textbf{H}_1{\bf{\Sigma}}_v\textbf{H}_1^{\dagger}\right)}\right).
\end{equation}

\subsection{Achievability Proof of (\ref{rate-pre-s})}\label{rdpcs}
This proof relies on the concept of backward decoding, which was introduced in \cite{Wil:InfoResuDiscMemo:Oct:82}.

\subsubsection{Block Markov Encoding and Backward Decoding}
Consider $B$+1 blocks of transmission, each consisting of $n$ symbols.  A sequence of $B$ messages, $w_i$ = ($w_{u,i}$, $w_{v,i}$) $\in$ $\mathcal{W}$, $i$ = 1,2,\ldots,$B$, each selected independently and uniformly over $\mathcal{W}$ is to be sent over the channel in $n(B+1)$ transmissions.

The senders use a triply-indexed set of codewords:
\begin{equation}
\begin{array}{lll}
\mathcal{C} & = & \{(v^n(w_v),u^n(k, w_{2u}),x_2^n(w_{2u})):w_v \in \{\phi,1,2,\ldots,2^{nR_v}\}, \\
& & k \in \{1,2,\ldots,2^{n(I(U;Y_1|X_2)-\delta(\epsilon))}\},w_{2u} \in \{\phi,1,2,\ldots,2^{nR_u}\}\}.
\end{array}
\end{equation}

$w_{2u}$ is sent cooperatively by both senders in block $i$ to help the receiver decode the previous message $w_{u,i-1}$.  To be more specific, the message $w_{2u}$ is the relay's estimate of the transmitter's message $w_u$ in the previous block.  See Table \ref{bac-dec} for details.

Backward decoding is employed at the receiver to decode $w_{u,i}$ and $w_{v,i}$.  Thus, after block $B$+1, $\textbf{y}(B+1)$ is used to decode $w_{u,B}$ and $w_{v,B}$.  Then, $\textbf{y}(B)$ and $w_{u,B}$ are used to decode $w_{u,B-1}$ and $w_{v,B-1}$.  Next, $\textbf{y}(B-1)$ and $w_{u,B-1}$ are used to decode $w_{u,B-2}$ and $w_{v,B-2}$.  The process continues until $\textbf{y}(2)$ and $w_{u,2}$ are used to decode $w_{u,1}$ and $w_{v,1}$.

\subsubsection{Generation of Random Code}
Fix $p(v)p(x_2)p(u|x_2)p(x_1|u,v)$.  Generate at random $2^{nR_v}$ i.i.d. $v^n$ sequences according to $\sim\prod_{i=1}^{n}p(v_i)$, and index them as $v^n(w_v),w_v\in\{1,2,\ldots,2^{nR_v}\}$.  Generate at random $2^{nR_u}$ i.i.d. $x_2^n$ sequences according to $\sim\prod_{i=1}^{n}p(x_{2i})$, and index them as $x_2^n(w_{2u}),w_{2u}\in\{1,2,\ldots,2^{nR_u}\}$.  For each $x_2^n(w_{2u})$, generate $2^{n(I(U;Y_1|X_2)-\epsilon)}$ conditionally independent $u^n\in A_{\epsilon}^{(n)}(u)$ sequences according to $p(u|x_2)$, and partition them into $2^{nR_u}$ equal-sized bins for each $x_2^n(w_{2u})$.  This defines the random codebook $\mathcal{C} = \{(v^n(w_v),u^n(k,w_{2u}),x_2^n(w_{2u}))\}$.  Finally, if $(u^n,v^n,x_2^n)\in A_{\epsilon}^{(n)}$, generate the codeword $x_1^n$ via $p(x_1^n|u^n,v^n)$.

The bin partitioning of the $u^n$ sequences implicitly defines a function $\mathcal{F}$ where $\mathcal{F}:u^n(k,w_{2u})\rightarrow w_u$.  Here, $k\in\{1,2,\ldots,2^{n(I(U;Y_1|X_2)-\epsilon)}\},w_{2u}\in\{\phi,1,2,\ldots,2^{nR_u}\}$, and $w_u\in\{\phi,1,2,\ldots,2^{nR_u}\}$.  For example, $\mathcal{F}(u^n(1,w_{2u})) = \mathcal{F}(u^n(2,w_{2u})) = \ldots = \mathcal{F}(u^n(2^{n(I(U;Y_1|X_2)-R_u-\epsilon)},w_{2u})) = w_u = 1$.  We see that $\mathcal{F}$ maps sequences $u^n(k,w_{2u})$ into their corresponding bin (and hence, message) indices $w_u$.  Since there is a one-to-one mapping between a sequence $u^n(k,w_{2u})$ and its bin $w_u$, we can also write $\mathcal{F}(u^n(k,w_{2u}))$ as $\mathcal{F}(k,w_{2u})$.

\subsubsection{Encoding}
Let $w_{u,i}\in\{1,2,\ldots,2^{nR_u}\}$ and $w_{v,i}\in\{1,2,\ldots,2^{nR_v}\}$ comprise the new message to be sent in block $i$.  Then, select any $u^n(k,w_{u,i-1})$ in bin $w_{u,i}$ such that $(x_2^n(w_{u,i-1}),u^n(k,w_{u,i-1}),v^n(w_{v,i}))\in A_{\epsilon}^{(n)}$ if this triplet exists.  Use the selected $u^n$ along with $v^n(w_{v,i})$ to generate $x_1^n$ via $p(x_1^n|u^n,v^n)$ and transmit this $x_1^n$.

Here, $P((x_1^n,u^n(k,w_{u,i-1}),v^n(w_{v,i}))\in A_{\epsilon}^{(n)}) > 1 - \epsilon$.

Assuming that the relay estimated $\hat{w}_{u,i-1}$ for $w_{u,i-1}$ in block $i$ - 1, then the relay sends $x_2^n(\hat{w}_{u,i-1})$ in block $i$.

\subsubsection{Encoding and Decoding Error Analysis}
We first perform an error event analysis for the encoding stage.

Encoding stage: The transmitter looks for a $u^n(k,w_{u,i-1})$ such that \\ $(u^n(k,w_{u,i-1}),v^n(w_{v,i}),x_2^n(w_{u,i-1}))\in A_{\epsilon}^{(n)}$.  If a sequence $u^n(k,w_{u,i-1})$ satisfying this criterion can be found, then the relay declares $\mathcal{F}(u^n(k,w_{u,i-1}))$ as $\hat{w}_{u,i}$.  Here, $E_{0i}$ is the event where
$(u^n,v^n(w_{v,i}),x_2^n(w_{u,i-1}))\notin A_{\epsilon}^{(n)}\forall u^n$ such that $\mathcal{F}(u^n) = w_{u,i}$; we have
\begin{eqnarray}
P(E_{0i}) & = & P((\textbf{u},\textbf{v},\textbf{x}_2(w_{u,i-1})) \notin A_{\epsilon}^{(n)}) \nonumber \\
& = & \sum_{v^n} p(v^n)P((\textbf{u},v^n,\textbf{x}_2(w_{u,i-1})) \notin A_{\epsilon}^{(n)}) \nonumber \\
& = & \sum_{v^n} p(v^n)(1-P((u^n,v^n,\textbf{x}_2(w_{u,i-1})) \in A_{\epsilon}^{(n)}))^{2^{n(I(U;Y_1|X_2)-R_u-\epsilon)}} \nonumber \\
& \leq & e^{-2^{n(I(U;Y_1|X_2)-R_u-\epsilon)}2^{-n(I(U;V|X_2)+\epsilon)}} \label{dpcsstep1}
\end{eqnarray}
which is arbitrarily small for $n$ sufficiently large if
\begin{equation}
R_u < I(U;Y_1|X_2) - I(U;V|X_2) - 2\epsilon.
\end{equation}

We note that (\ref{dpcsstep1}) follows from the following two facts: \\
i) $2^{-n(I(U;V|X_2)+\epsilon)}\leq P((u^n,v^n,\textbf{x}_2(w_{u,i-1}))\in A_{\epsilon}^{(n)})$ and \\
ii) $(1-x)^k\leq e^{-kx}$ for $0\leq x\leq 1$ and $k\geq 1$.

Thus, $\hat{w}_{u,i}$ = $w_{u,i}$ with $P(E_{0i})$ arbitrarily small if $n$ is sufficiently large and if
\begin{equation} \label{rdpcs1}
R_u < I(U;Y_1|X_2) - I(U;V|X_2).
\end{equation}

Note that for $w_{u,i}$ and $w_{v,i}$, we perform backward decoding at the receiver, though we perform block-by-block decoding at the relay.  The following analysis is for the case where the receiver attempts to decode $w_u$ before decoding $w_v$.

Here, we proceed through three decoding steps.  We employ the concept of weak typicality.  Define the following error events:
\begin{itemize}
\item $E_{1i}$ as the event that $(u^n(k_i,w_{2u,i}),x_2^n(w_{2u,i}),y_1^n(i),y^n(i))\notin A_{\epsilon}^{(n)}$, where $y_1^n(i)$ and $y^n(i)$ are the observations by the relay and receiver, respectively in block $i$.
\item $E_{mi}$ as the event that there is an error in block $i$ at decoding step $m-1$ for $m$ = 2,3,4.
\end{itemize}

Thus, the overall probability of error $P_e^{(n)} = P(\bigcup_{m=0}^{4} E_{mi})\leq\sum_{m=0}^{4} P(E_{mi})$.  We first note that for $n$ sufficiently large, $P(E_{1i}) < \epsilon$ by the asymptotic equipartition property (AEP).  Now we bound $P(E_{mi})$ for $m$ = 2,3,4 as follows.

Decoding step 1: Upon observing $y_1^n(i)$, the relay receiver declares that $\hat{w}_{u,i} = \hat{w}_u$ is sent if it is the unique index such that $(u^n(\hat{k}_i,w_{u,i-1}),x_2^n(w_{u,i-1}),y_1^n(i))\in A_{\epsilon}^{(n)}$, where $u^n(\hat{k}_i,w_{u,i-1})$ is in bin $\hat{w}_{u,i}$.  Here, $E_{2i}$ is the event that $\exists\hat{w}_u\neq w_{u,i}$ such that $(u^n(\hat{k}_i,w_{u,i-1}),x_2^n(w_{u,i-1}),y_1^n(i))\in T_{\epsilon}^{(n)}$, where $u^n(\hat{k}_i,w_{u,i-1})$ is in bin $\hat{w}_u$.  Now, for $\hat{w}_u$ $\neq$ $w_{u,i}$,
\begin{eqnarray}
P(E_{2i}|\hat{w}_u) & = & P((\textbf{u}(\hat{k}_i,w_{u,i-1}),\textbf{x}_2(w_{u,i-1}),\textbf{y}_1(i))\in A_{\epsilon}^{(n)}) \nonumber \\
& = & \sum_{(u^n(\hat{k}_i,w_{u,i-1}),y_1^n(i))\in A_{\epsilon}^{(n)}(U,Y_1|x_2^n),\hat{w}_u\neq w_{u,i}} p(u^n(\hat{k}_i,w_{u,i-1}),y_1^n(i)|x_2^n(w_{u,i-1}) \nonumber \\
& = & \sum_{(u^n(\hat{k}_i,w_{u,i-1}),y_1^n(i))\in A_{\epsilon}^{(n)}(U,Y_1|x_2^n),\hat{w}_u\neq w_{u,i}} p(u^n(\hat{k}_i,w_{u,i-1})|x_2^n(w_{u,i-1}))p(y_1^n(i)|x_2^n(w_{u,i-1})) \label{supsstep1} \\
& \leq & \left|A_{\epsilon}^{(n)}(U,Y_1|x_2^{n})\right| 2^{-n(H(U|X_2)-\epsilon)}2^{-n(H(Y_1|X_2)-\epsilon)} \nonumber \\
& \leq & 2^{-n(H(U|X_2)+H(Y_1|X_2)-H(U,Y_1|X_2)-3\epsilon)} \nonumber \\
& = & 2^{-n(H(Y_1|X_2)-H(Y_1|X_2,U)-3\epsilon)} \nonumber \\
& = & 2^{-n(I(U;Y_1|X_2)-3\epsilon)} \nonumber
\end{eqnarray}
where (\ref{supsstep1}) follows from the fact that $\textbf{y}_1(i)$ and $\textbf{u}(\hat{k}_i,w_{u,i-1})$ are independent for $\hat{w}_u\neq w_{u,i}$.  Thus, we have
\begin{equation}
P(E_{2i}) \leq 2^{nR_u} \cdot 2^{-n(I(U;Y_1|X_2)-3\epsilon)}
\end{equation}
and so $\hat{w}_{u,i} = w_{u,i}$ with $P(E_{2i})$ arbitrarily small if $n$ is sufficiently large and if
\begin{equation} \label{rsups1}
R_u < I(U;Y_1|X_2).
\end{equation}

Decoding step 2: Backward decoding is employed to estimate $w_{u,i}$ at the receiver.  Assume that the receiver has estimated $\tilde{w}_{u,i+1}$ for $w_{u,i+1}$.  Now, the receiver looks for a unique $w_{2u}$ such that \\ 
($u^n(k_i,w_{2u}),x_2^n(w_{2u}),y^n(i))\in A_{\epsilon}^{(n)}$, where $\mathcal{F}(k_i,w_{2u}) = \tilde{w}_{u,i+1}$.  It then declares $\tilde{w}_u = w_{2u}$ if this unique $w_{2u}$ exists.  Here, $E_{3i}$ is the event that $\exists\tilde{w}_u\neq w_{u,i}$ such that $(u^n(k_i,\tilde{w}_u),x_2^n(\tilde{w}_u),y^n(i))\in A_{\epsilon}^{(n)}$, where $\mathcal{F}(k_i,\tilde{w}_u) = \tilde{w}_{u,i+1}$.  Now, for $\tilde{w}_u\neq w_{u,i}$,
\begin{eqnarray}
P(E_{3i}|\tilde{w}_u) & = & P((\textbf{u},\textbf{x}_2(\tilde{w}_u),\textbf{y}(i)) \in A_{\epsilon}^{(n)}) \nonumber \\
& = & \sum_{(u^n(k_i,\tilde{w}_u),x_2^n(\tilde{w}_u),y^n(i)) \in A_{\epsilon}^{(n)}(U,X_2,Y),\tilde{w}_u\neq w_{u,i},\mathcal{F}(k_i,\tilde{w}_u) = \tilde{w}_{u,i+1}} \\
& & p(u^n(k_i,\tilde{w}_u),x_2^n(\tilde{w}_u),y^n(i)) \nonumber \\
& = & \sum_{(u^n(k_i,\tilde{w}_u),x_2^n(\tilde{w}_u),y^n(i)) \in A_{\epsilon}^{(n)}(U,X_2,Y),\tilde{w}_u\neq w_{u,i},\mathcal{F}(k_i,\tilde{w}_u) = \tilde{w}_{u,i+1}} \nonumber \\
& & p(u^n(k_i,\tilde{w}_u),x_2^n(\tilde{w}_u))p(y^n(i)) \label{dpcsstep2} \\
& \leq & \left|A_{\epsilon}^{(n)}(U,X_2,Y)\right|2^{-n(H(U,X_2)-\epsilon)}2^{-n(H(Y)-\epsilon)} \nonumber \\
& \leq & 2^{-n(H(U,X_2)+H(Y)-H(U,X_2,Y)-3\epsilon)} \nonumber \\
& = & 2^{-n(I(U,X_2;Y)-3\epsilon)} \nonumber
\end{eqnarray}
where (\ref{dpcsstep2}) follows from the fact that $\textbf{y}(i)$ and $(\textbf{u}(k_i,\tilde{w}_u),\textbf{x}_2(\tilde{w}_u))$ are independent for $\tilde{w}_u\neq w_{u,i}$.  Thus, we have
\begin{equation}
P(E_{3i}) \leq 2^{nR_u} \cdot 2^{-n(I(U,X_2;Y)-3\epsilon)}
\end{equation}
and so $\tilde{w}_u$ = $w_{u,i}$ with $P(E_{3i})$ arbitrarily small if $n$ is sufficiently large and if
\begin{equation} \label{rdpcs2}
R_u < I(U,X_2;Y).
\end{equation}

Now, we combine (\ref{rdpcs1}), (\ref{rsups1}) and (\ref{rdpcs2}) to obtain
\begin{equation}
R_u < \min((I(U;Y_1|X_2)-I(U;V|X_2)),I(U,X_2;Y)).
\end{equation}

Decoding step 3: Backward decoding is also employed to estimate $w_{v,i}$ at the receiver.  Assume that the receiver has estimated $\tilde{w}_{u,i+1}$ for $w_{u,i+1}$.  Recall that the receiver has estimated $\tilde{w}_{u,i}$ for $w_{u,i}$ in decoding step 2.  Now, the receiver looks for a unique $w_v$ such that $(u^n(k_i,\tilde{w}_{u,i}),x_2^n(\tilde{w}_{u,i}),y^n(i),v^n(w_v))\in A_{\epsilon}^{(n)}$, where $\mathcal{F}(k_i,\tilde{w}_{u,i}) = \tilde{w}_{u,i+1}$.  It then declares $\tilde{w}_v = w_v$ if this unique $w_v$ exists.  Here, $E_{4i}$ is the event that $\exists\tilde{w}_v\neq w_{v,i}$ such that $(u^n(k_i,\tilde{w}_{u,i}),x_2^n(\tilde{w}_{u,i}),y^n(i),v^n(\tilde{w}_v))\in A_{\epsilon}^{(n)}$, where $\mathcal{F}(k_i,\tilde{w}_{u,i}) = \tilde{w}_{u,i+1}$.  Now, for $\tilde{w}_v\neq w_{v,i}$,
\begin{eqnarray}
P(E_{4i}|\tilde{w}_v) & = & P((\textbf{u},\textbf{x}_2(\tilde{w}_{u,i}),\textbf{y}(i),\textbf{v}(\tilde{w}_v)) \in A_{\epsilon}^{(n)}) \nonumber \\
& = & \sum_{(v^n(\tilde{w}_v),y^n(i)) \in A_{\epsilon}^{(n)}(V,Y|u^n,x_2^n),\tilde{w}_v\neq w_{v,i},\mathcal{F}(k_i,\tilde{w}_{u,i}) = \tilde{w}_{u,i+1}} \nonumber \\ & & p(v^n(\tilde{w}_v),y^n(i)|u^n(k_i,\tilde{w}_{u,i}),x_2^n(\tilde{w}_{u,i})) \nonumber \\
& = & \sum_{(v^n(\tilde{w}_v),y^n(i)) \in A_{\epsilon}^{(n)}(V,Y|u^n,x_2^n),\tilde{w}_v\neq w_{v,i},\mathcal{F}(k_i,\tilde{w}_{u,i}) = \tilde{w}_{u,i+1}} \nonumber \\ & & p(v^n(\tilde{w}_v)|u^n(k_i,\tilde{w}_{u,i}),x_2^n(\tilde{w}_{u,i}))p(y^n(i)|u^n(k_i,\tilde{w}_{u,i}),x_2^n(\tilde{w}_{u,i})) \label{dpcsstep3} \\
& \leq & \left|A_{\epsilon}^{(n)}(V,Y|u^n,x_2^n)\right|2^{-n(H(V|U,X_2)-\epsilon)}2^{-n(H(Y|U,X_2)-\epsilon)} \nonumber \\
& \leq & 2^{-n(H(V|U,X_2)+H(Y|U,X_2)-H(V,Y|U,X_2)-3\epsilon)} \nonumber \\
& = & 2^{-n(I(V;Y|U,X_2)-3\epsilon)} \nonumber
\end{eqnarray}
where (\ref{dpcsstep3}) follows from the fact that $\textbf{y}(i)$ and $\textbf{v}(\tilde{w}_v)$ are independent for $\tilde{w}_v\neq w_{v,i}$.  Thus, we have
\begin{equation}
P(E_{4i}) \leq 2^{nR_v} \cdot 2^{-n(I(V;Y|U,X_2)-3\epsilon)}
\end{equation}
and so $\tilde{w}_v = w_{v,i}$ with $P(E_{4i})$ arbitrarily small if $n$ is sufficiently large and if
\begin{equation}
R_v < I(V;Y|U,X_2).
\end{equation}

\subsection{Achievability Proof of (\ref{rate-pre-c})}\label{rdpcc}
This proof also relies on the concept of backward decoding.  Apply the code generation and encoding procedures from Section \ref{rdpcs}.  Note that in this case, backward decoding is employed at the receiver to decode $w_{u,i}$, not both $w_{u,i}$ and $w_{v,i}$.  Thus, after block $B$+1, $\textbf{y}(B+1)$ is used to decode $w_{u,B}$.  Then, $\textbf{y}(B)$ and $w_{u,B}$ are used to decode $w_{u,B-1}$.  Next, $\textbf{y}(B-1)$ and $w_{u,B-1}$ are used to decode $w_{u,B-2}$.  The process continues until $\textbf{y}(2)$ and $w_{u,2}$ are used to decode $w_{u,1}$.  The receiver can use block-by-block decoding to decode $w_{v,i}$; thus, $w_{v,i}$ can be decoded in block $i$ after $\textbf{y}(i)$ is received, where $i$ = 1,2,\ldots,$B$.

\subsubsection{Encoding and Decoding Error Analysis}
We first perform an error event analysis for the encoding stage.

Encoding stage: The analysis for this stage is similar to the analysis for the encoding stage in Section \ref{rdpcs}.  Thus we have
\begin{equation} \label{rdpcc1}
R_u < I(U;Y_1|X_2) - I(U;V|X_2).
\end{equation}

Note that for $w_{u,i}$, we perform backward decoding at the receiver, though we still perform block-by-block decoding at the relay.  We also perform block-by-block decoding at the receiver for $w_{v,i}$.

Once again, we proceed through three decoding steps and employ the concept of weak typicality.  Define the following error events:
\begin{itemize}
\item $E_{1i}$ as the event that $(u^n(k_i,w_{2u,i}),x_2^n(w_{2u,i}),y_1^n(i),y^n(i))\notin A_{\epsilon}^{(n)}$, where $y_1^n(i)$ and $y^n(i)$ are the observations by the relay and receiver, respectively in block $i$.
\item $E_{mi}$ as the event that there is an error in block $i$ at decoding step $m-1$ for $m$ = 2,3,4.
\end{itemize}

Thus, the overall probability of error $P_e^{(n)} = P(\bigcup_{m=1}^{4} E_{mi})\leq\sum_{m=1}^{4} P(E_{mi})$.  We first note that for $n$ sufficiently large, $P(E_{1i}) < \epsilon$ by the asymptotic equipartition property (AEP).  Now we bound $P(E_{mi})$ for $m$ = 2,3,4 as follows.

Decoding step 1: Upon observing $y^n(i)$, the receiver declares that $\tilde{w}_{v,i} = \tilde{w}_v$ is sent if it is the unique index such that $(v^n(\tilde{w}_v),y^n(i))\in A_{\epsilon}^{(n)}$.  Here, $E_{2i}$ is the event that $\exists\tilde{w}_v\neq w_{v,i}$ such that $(v^n(\tilde{w}_v),y^n(i))\in A_{\epsilon}^{(n)}$.  Now, for $\tilde{w}_v\neq w_{v,i}$,
\begin{eqnarray}
P(E_{2i}|\tilde{w}_v) & = & P((\textbf{v}(\tilde{w}_v),\textbf{y}(i)) \in A_{\epsilon}^{(n)}) \nonumber \\
& = & \sum_{(v^n(\tilde{w}_v),y^n(i)) \in A_{\epsilon}^{(n)}(V,Y),\tilde{w}_v\neq w_{v,i}} p(v^n(\tilde{w}_v),y^n(i)) \nonumber \\
& = & \sum_{(v^n(\tilde{w}_v),y^n(i)) \in A_{\epsilon}^{(n)}(V,Y),\tilde{w}_v\neq w_{v,i}} p(v^n(\tilde{w}_v))p(y^n(i)) \label{dpccstep1} \\
& \leq & \left|A_{\epsilon}^{(n)}(V,Y)\right| 2^{-n(H(V)-\epsilon)}2^{-n(H(Y)-\epsilon)} \nonumber \\
& \leq & 2^{-n(H(V)+H(Y)-H(V,Y)-3\epsilon)} \nonumber \\
& = & 2^{-n(I(V;Y)-3\epsilon)} \nonumber
\end{eqnarray}
where (\ref{dpccstep1}) follows from the fact that $\textbf{y}(i)$ and $\textbf{v}(\tilde{w}_v)$ are independent for $\tilde{w}_v\neq w_{v,i}$.  Thus, we have
\begin{equation}
P(E_{2i}) \leq 2^{nR_v} \cdot 2^{-n(I(V;Y)-3\epsilon)}
\end{equation}
and so $\tilde{w}_{v,i} = w_{v,i}$ with $P(E_{2i})$ arbitrarily small if $n$ is sufficiently large and if
\begin{equation}
R_v < I(V;Y).
\end{equation}

Decoding step 2: The analysis for this step is similar to the analysis for decoding step 1 in Section \ref{rdpcs}.  Thus we have
\begin{equation} \label{rsupc1}
R_u < I(U;Y_1|X_2).
\end{equation}



Decoding step 3: Backward decoding is employed to estimate $w_{u,i}$ at the receiver.  Assume that the receiver has estimated $\tilde{w}_{u,i+1}$ for $w_{u,i+1}$.  Recall that the receiver has estimated $\tilde{w}_{v,i}$ for $w_{v,i}$ in decoding step 1.  Now, the receiver looks for a unique $w_{2u}$ such that $(u^n(k_i,w_{2u}),x_2^n(w_{2u}),y^n(i),v^n(\tilde{w}_{v,i}))\in A_{\epsilon}^{(n)}$, where $\mathcal{F}(k_i,w_{2u}) = \tilde{w}_{u,i+1}$.  It then declares $\tilde{w}_u = w_{2u}$ if this unique $w_{2u}$ exists.  Here, $E_{4i}$ is the event that $\exists\tilde{w}_u\neq w_{u,i}$ such that $(u^n(k_i,\tilde{w}_u),x_2^n(\tilde{w}_u),y^n(i),v^n(\tilde{w}_{v,i}))\in A_{\epsilon}^{(n)}$, where $\mathcal{F}(k_i,\tilde{w}_u) = \tilde{w}_{u,i+1}$.  Now, for $\tilde{w}_u\neq w_{u,i}$,
\begin{eqnarray}
P(E_{4i}|\tilde{w}_u) & = & P((\textbf{u},\textbf{x}_2(\tilde{w}_u),\textbf{y}(i), \textbf{v}(\tilde{w}_{v,i})) \in A_{\epsilon}^{(n)}) \nonumber \\
& = & \sum_{(u^n(k_i,\tilde{w}_u),x_2^n(\tilde{w}_u),y^n(i)) \in A_{\epsilon}^{(n)}(U,X_2,Y|v^n),\tilde{w}_u\neq w_{u,i},\mathcal{F}(k_i,\tilde{w}_u) = \tilde{w}_{u,i+1}} \nonumber \\ & & p(u^n(k_i,\tilde{w}_u),x_2^n(\tilde{w}_u),y^n(i)|v^n(\tilde{w}_{v,i})) \nonumber \\
& = & \sum_{(u^n(k_i,\tilde{w}_u),x_2^n(\tilde{w}_u),y^n(i)) \in A_{\epsilon}^{(n)}(U,X_2,Y|v^n),\tilde{w}_u\neq w_{u,i},\mathcal{F}(k_i,\tilde{w}_u) = \tilde{w}_{u,i+1}} \nonumber \\ & & p(u^n(k_i,\tilde{w}_u),x_2^n(\tilde{w}_u)|v^n(\tilde{w}_{v,i})p(y^n(i)|v^n(\tilde{w}_{v,i}) \label{dpccstep3} \\
& \leq & \left|A_{\epsilon}^{(n)}(U,X_2,Y|v^n)\right| 2^{-n(H(U,X_2|V)-\epsilon)}2^{-n(H(Y|V)-\epsilon)} \nonumber \\
& \leq & 2^{-n(H(U,X_2|V)+H(Y|V)-H(U,X_2,Y|V)-3\epsilon)} \nonumber \\
& = & 2^{-n(I(U,X_2;Y|V)-3\epsilon)} \nonumber
\end{eqnarray}
where (\ref{dpccstep3}) follows from the fact that $\textbf{y}(i)$ and ($\textbf{u}(k_i,\tilde{w}_u)$,$\textbf{x}_2(\tilde{w}_u)$) are independent for $\tilde{w}_u$ $\neq$ $w_{u,i}$.  Thus, we have
\begin{equation}
P(E_{4i}) \leq 2^{nR_u} \cdot 2^{-n(I(U,X_2;Y|V)-3\epsilon)}
\end{equation}
and so $\tilde{w}_u = w_{u,i}$ with $P(E_{4i})$ arbitrarily small if $n$ is sufficiently large and if
\begin{equation} \label{rdpcc2}
R_u < I(U,X_2;Y|V).
\end{equation}

Now, we combine (\ref{rdpcc1}), (\ref{rsupc1}) and (\ref{rdpcc2}) to obtain
\begin{equation}
R_u < \min((I(U;Y_1|X_2)-I(U;V|X_2)),I(U,X_2;Y|V)).
\end{equation}

\begin{table*}[tb]
\begin{center}
\caption{Block Markov encoding and backward decoding}
\begin{tabular}{c|ccccc}
Block&1&2&\ldots&B&B+1\\
\hline
$\mathcal{V}$&$v^n(w_{v,1})$&$v^n(w_{v,2})$&\ldots&$v^n(w_{v,B})$&$v^n(\phi^{'})$\\
$\mathcal{U}$&$u^n(k_1,\phi)$&$u^n(k_2,w_{u,1})$&\ldots&$u^n(k_B,w_{u,B-1})$&$u^n(k_{B+1},w_{u,B})$\\  $X_2$&$x_2^n(\phi)$& $x_2^n(w_{u,1})$&\ldots&$x_2^n(w_{u,B-1})$&$x_2^n(w_{u,B})$\\
$Y_1$&$\hat{w}_{u,1}$&$\hat{w}_{u,2}$&\ldots&$\hat{w}_{u,B}$&$\phi^{''}$
\end{tabular}
\label{bac-dec}
\end{center}
\end{table*}

\end{document}